\documentclass[english,prb,amsmath,amsfonts,superscriptaddress, twocolumn,nofootinbib]{revtex4-2}
\usepackage[LGR,T1]{fontenc}
\usepackage[latin9]{inputenc}
\usepackage{array}
\usepackage{verbatim}
\usepackage{units}
\usepackage{textcomp}
\usepackage{multirow}
\usepackage{amsmath}
\usepackage{amssymb}
\usepackage{graphicx}

\makeatletter

\DeclareRobustCommand{\greektext}{%
  \fontencoding{LGR}\selectfont\def\encodingdefault{LGR}}
\DeclareRobustCommand{\textgreek}[1]{\leavevmode{\greektext #1}}
\ProvideTextCommand{\~}{LGR}[1]{\char126#1}

\newcommand{\lyxmathsym}[1]{\ifmmode\begingroup\def\b@ld{bold}
  \text{\ifx\math@version\b@ld\bfseries\fi#1}\endgroup\else#1\fi}

\providecommand{\tabularnewline}{\\}

\makeatother

\usepackage{babel}
\begin{document}
\title{Ultrafast optical polarimetry in magnetic phases of Kondo semi metal
CeSb}
\author{M. Naseska}
\affiliation{Department of Complex Matter, Jozef Stefan Institute, Jamova 39, 1000
Ljubljana, Slovenia}
\author{N. D. Zhigadlo}
\affiliation{CrystMat Company, CH-8037 Zurich, Switzerland}
\author{Z. Jagli\v{c}i\'{c}}
\affiliation{Faculty of Civil and Geodetic Engineering, University of Ljubljana,
Jamova cesta 2, Ljubljana, Slovenia}
\affiliation{Institute of Mathematics, Physics and Mechanics, Jadranska 19, Ljubljana,
Slovenia}
\author{E. Goreshnik}
\affiliation{Dept. of Inorganic Chemistry and Technology, Jozef Stefan Institute,
Jamova 39, 1000 Ljubljana, Slovenia}
\author{T. Mertelj}
\email{tomaz.mertelj@ijs.si}

\affiliation{Department of Complex Matter, Jozef Stefan Institute, Jamova 39, 1000
Ljubljana, Slovenia}
\affiliation{Center of Excellence for Nanoscience and Nanotechnology (CENN Nanocenter),
Jamova 39, 1000 Ljubljana, Slovenia}
\date{\today}
\begin{abstract}
We investigated photoinduced ultrafast transient dynamics in different
magnetic phases in CeSb by means of the time resolved magneto-optical
spectroscopy. We observe a distinctive coherent oscillations in the
ground-state antiferromagnetic (AF) phase and the high-magnetic field
ferromagnetic (F) phase. While the AF-phase oscillations frequencies
match the recent Raman scattering findings the F-phase oscillation
frequency does not correspond to the previously observed magnetic
excitation. The large spectroscopic factor, $g=3.94$, and optical
polarization properties suggest that it corresponds to a previously
undetected Ce$^{3+}$ coherent crystal-field state excitation. The
AF-phase oscillations show no magnetic field dependence so their lattice
origin cannot be entirely excluded. The non-oscillatory part of the
transients is qualitatively similar in all investigated magnetic phases
with a faster sub-picosecond dynamics in the ferromagnetic and ferro-para-magnetic
phases and is attributed to differences in the electronic structure,
which affect the photo-excited quasiparticle energy relaxation kinetics.
\end{abstract}
\maketitle

\section{Introduction}

CeSb is a material that has one of the most complex phase diagrams
among lanthanide monopnictides. It contains at least 16 different
magnetic phases in the $H$-$T$ plane \citep{rossat-mignod1977phasediagram}
comprising different sequences of ferromagnetic, with either up or
down spin orientation, and paramagnetic (001) planes stacked along
the $c$-axis.\citep{rossat-mignod1977phasediagram,rossat-mignod1985neutron}
The complexity is thought to arise from the interplay of Kondo, spin-orbit
and crystal-field effects as well as the Sb-$5p$ and Ce-$4d$ orbital
mixing. \citep{halg1986anisotropic,jang2019directvisualization,kasuya1990themagnetic}
Lattice modulation in the magnetic phases was also observed \citep{mcmorrow1997xrayscattering}.

\begin{figure}
\includegraphics[width=1\columnwidth]{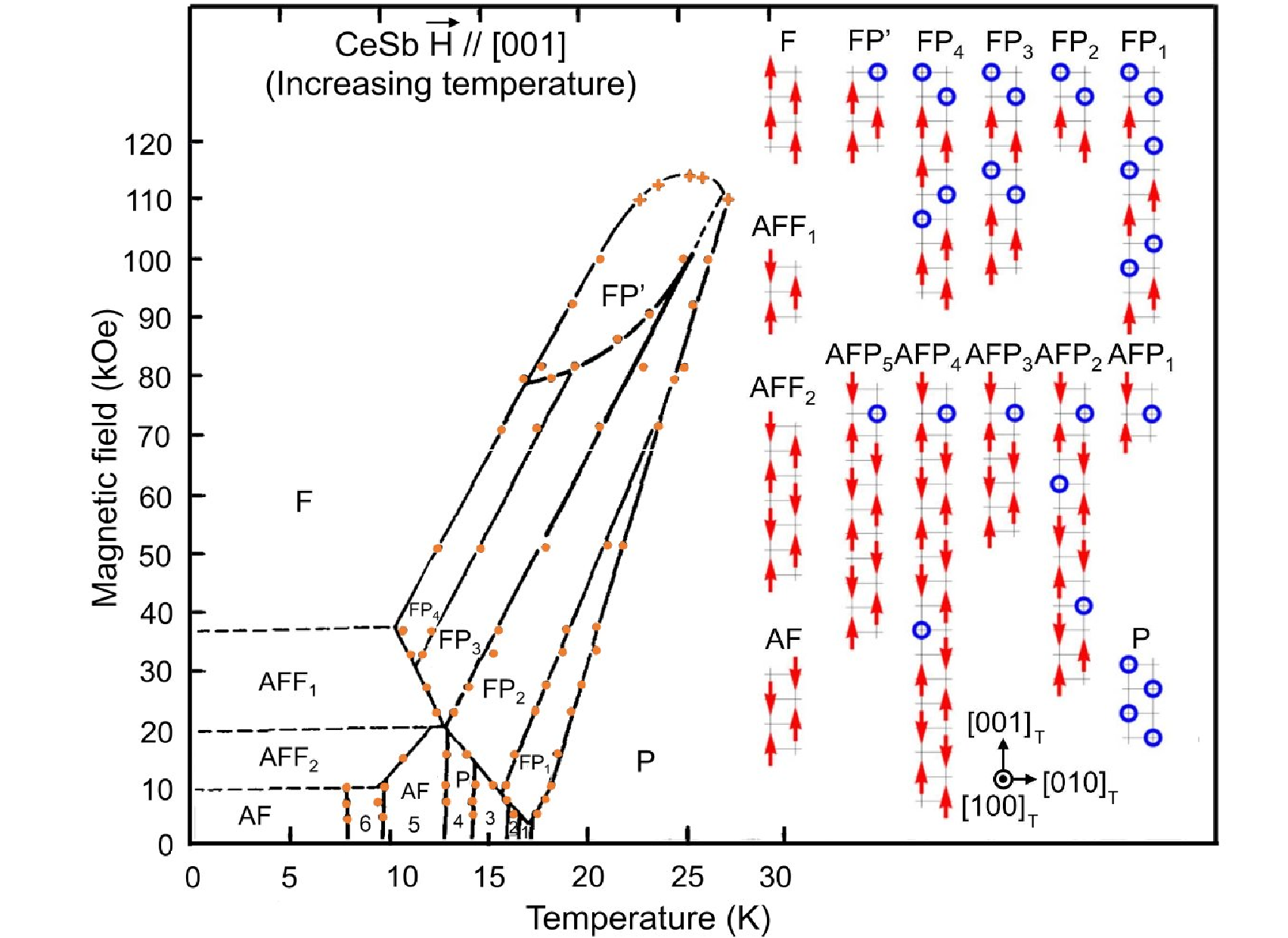}\caption{Magnetic phase diagram of CeSb. Adapted from \citet{rossat-mignod1985neutron}
and \citet{ye2018extreme}. Each red arrow represents a single ferromagnetically
ordered (001) plane while the blue circles represent the paramagnetic
(001) planes. The neighboring planes are due to the face centered
cubic structure shifted relatively by {[}0$\frac{1}{2}$$\frac{1}{2}${]}
(or {[}$\frac{1}{2}0$$\frac{1}{2}${]}). \label{fig:PHDia}}
\end{figure}

The phase diagram \citep{rossat-mignod1977phasediagram,rossat-mignod1985neutron}
(see\ Fig. \ref{fig:PHDia}) and some magnetic excitations \citep{rossat-mignod1985magnetic,halg1986anisotropic}
were thoroughly studied by neutron scattering. Recently the sensitivity
of the electronic structure to the magnetic phase has been demonstrated
\citep{kuroda2020devilsstaircase} and additional magnetic excitations
were found in the ground state AF phase \citep{arai2022multipole}.
While the main features of the magnetic behavior are understood and
successfully modeled using effective interaction approach \citep{kasuya1990themagnetic}
the microscopic origin of the interactions is still puzzling \citep{jang2019directvisualization,arai2022multipole}.
An insight into non-equilibrium dynamics of different phases might
therefore shed some light on to the interplay between different degrees
of freedom.

Here we present and discuss our investigation of the ultrafast non-equilibrium
dynamics upon photo excitation in different magnetic phases in CeSb
with focus on the magnetic excitations in the weakly non equilibrium
photo excited state. We confirm the presence of the recently reported
\citep{arai2022multipole} additional modes in the ground-state AF
phase and show that their frequencies are magnetic-field independent.
In the high-magnetic field F phase we identify a previously unobserved
$^{2}\overline{E}_{2\mathrm{g}}$ Ce$^{3+}$crystal field excited
state. The associated coherent oscillatory optical response can be
linked to the \textit{real-time quantum evolution} of the superposition
state involving the ground state and a particular excited crystal-field
state.

\section{Experimental}

\begin{figure}
\hfill{}\includegraphics[width=1\columnwidth]{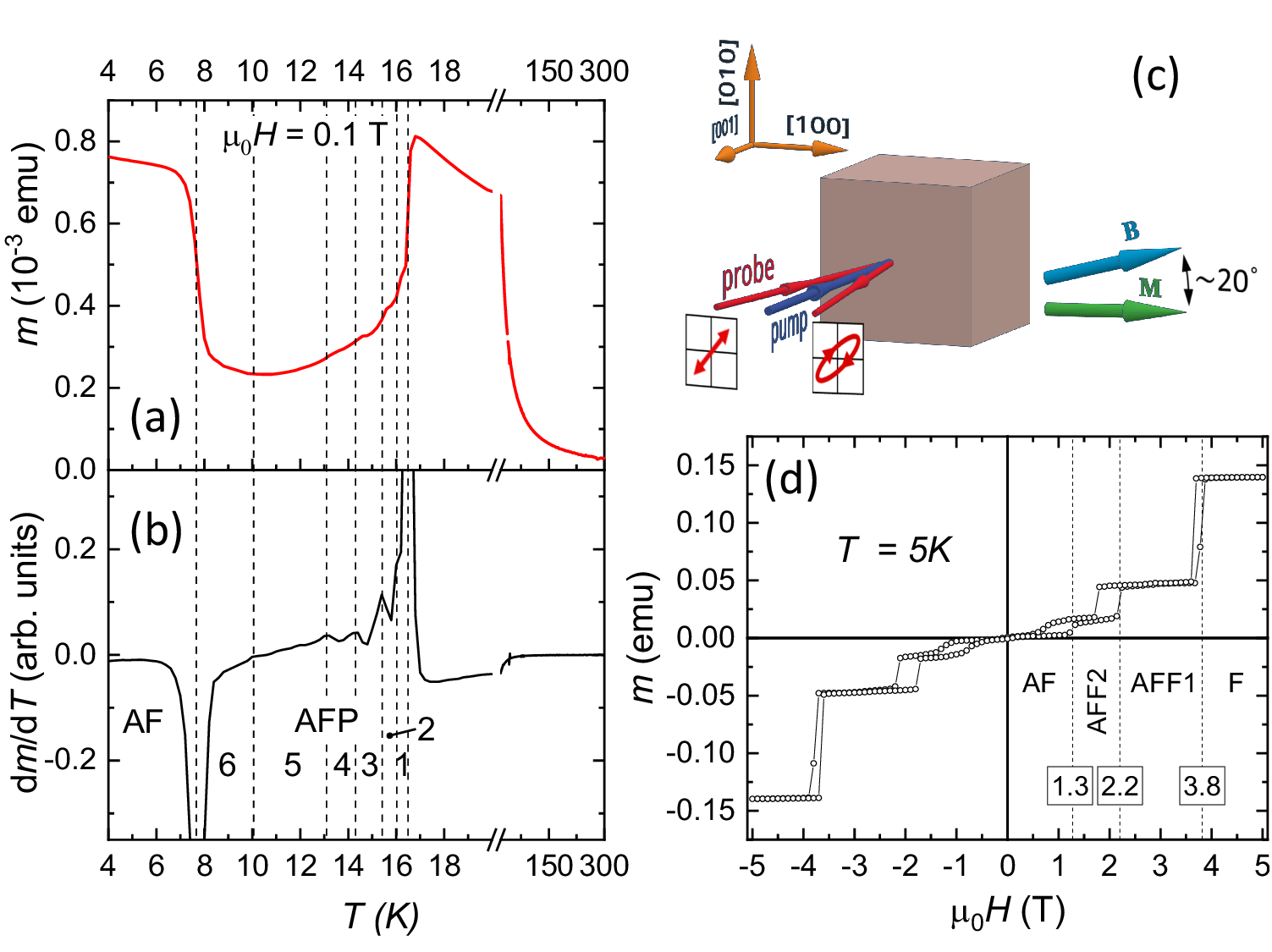}\hfill{}

\caption{Sample characterization data and the optical setup geometry. (a) The
low-field magnetic moment of the sample as a function of $T$ and
(b) the corresponding derivative. (c) The time resolved polarimetry
setup geometry. (d) The low-$T$ magnetic moment as a function of
magnetic field. \label{fig:figPPExp}}
\end{figure}

\subsection{Samples}

CeSb is the only congruently melting compound in the Ce-Sb phase diagram
with a relatively high melting temperature ($T_{\mathrm{m}}$ = 1760
°C), and its crystals can be grown by simply cooling a stoichiometric
mixture of Ce and Sb, or out of a Sn-rich flux \citep{canfield2019newmaterials}.
The present samples were produced by two-step process. High-purity
Ce (4N) and Sb (4N) were pre-reacted at 650 °C in evacuated quartz
ampule in the first step. The resultant powder was then compressed
into a pellet and put inside a molybdenum crucible. The sealed crucible
was heated to a temperature above the melting point of CeSb, then
slowly cooled to a temperature about 40 °C below the melting temperature,
maintained there for about a week, and finally cooled to room temperature.
Nicely formed single crystals up to 3 \texttimes{} 3 \texttimes{}
3 mm$^{3}$ in volume were collected. The 1:1 stoichiometry and excellent
structural quality of the produced crystals were confirmed by energy
dispersive X-ray spectroscopy and X-ray diffraction studies.

The single crystal used for the optical measurements was also characterized
by SQUID magnetometry. The low-field $T$-scan and low-$T$ field
magnetization scans shown in Fig. \ref{fig:figPPExp} (b) and (d)
were found consistent with the published phase diagram \citep{rossat-mignod1985neutron,wiener2000magnetic}
(Fig. \ref{fig:PHDia}).

\subsection{Optical measurements}

The sample was mounted in a 7-T optical split-coil cryomagnet with
transverse optical access. The pump (3.1-eV photon energy) and probe
(1.55-eV photon energy) pulse-train beams (pulse length $<100$ fs,
250 kHz repetition rate) were focused in a nearly perpendicular manner
on a {[}001{]} cleaved crystal facet {[}see Fig. \ref{fig:figPPExp}
(c){]}. The magnetic field was lying in the facet plane at $\sim20{^\circ}$
with respect to the {[}100{]} cubic direction, while the probe polarization
was oriented approximately along the {[}110{]} direction (unless specified
differently) to maximize the transient polarization rotation.

The beam diameters of slightly elliptical beams were of similar size.
The probe beam diameter was (64 x 52) \textmu m$^{2}$ and the pump
beam diameter was (80 x 76) \textmu m$^{2}$. The fluence of the probe
in all measurements was around 4 \textmu J/cm$^{2}$ and the pump
beam fluence was in the interval between 60-90 \textmu J/cm$^{2}$
(unless specified differently).

The reflected probe beam transient polarization rotation, $\Delta\Phi$,
was acquired using a Wollaston prism and a pair of balanced silicon
PIN photo diodes using the standard lock-in techniques. The scattered
pump contribution was suppressed by means of a long-pass optical filter.
The detector was kept carefully balanced to avoid contamination of
the $\Delta\Phi$ signal with the isotropic part of the transient
reflectivity, $\Delta R/R$, response. To measure $\Delta R/R$ the
detector was rotated to maximize the DC current component on one of
the PIN diodes while the other PIN diode was disconnected.

Since zero field cooling resulted in multi-domain states and non-reproducible
transients the sample was always field cooled at $\mu_{0}H=7$ T in
order to achieve a mono-domain magnetic state \citep{rossat-mignod1977phasediagram}.
Due to the field cooling the {[}001{]}$_{\mathrm{T}}$ tetragonal
direction and the spin orientations are along the {[}100{]} pseudo-cubic
direction \citep{rossat-mignod1977phasediagram} in the F, FP$_{n}$
and AFF$_{n}$ phases (see Fig. \ref{fig:PHDia} for designation of
the phases). In the ground-state AF phase we observed an angular momentum
flop (discussed later) resulting in the {[}001{]}$_{\mathrm{T}}$
direction and spin directions rotation to the {[}010{]} pseudo-cubic
direction. In the rest of the text we use the pseudocubic notation
tacitly assuming the above orientations.

\section{Results}

\begin{figure*}
\hfill{}\includegraphics[width=0.6\columnwidth]{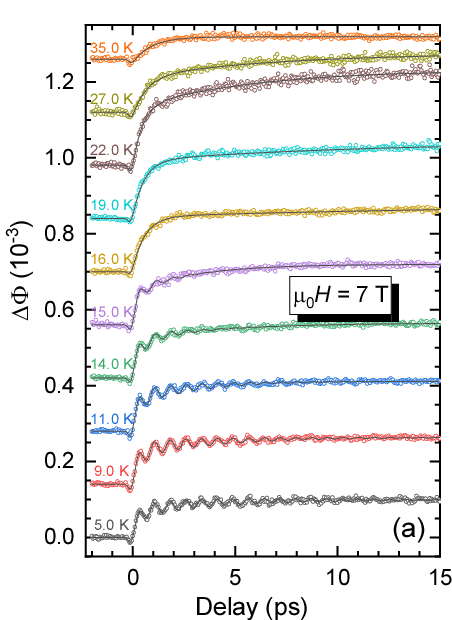}\hspace*{\fill}\includegraphics[width=0.6\columnwidth]{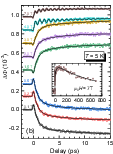}\hspace*{\fill}\caption{(a) Transient polarization rotation at selected $T$ at $\mu_{0}H=7$~T
(measured on warming) and (b) at selected $\mu_{0}H$ at $T=5\,$K
(measured on decreasing field). The traces are vertically shifted
for clarity. A typical longer timescale evolution of the signal is
shown in the inset to (b). The lines correspond to the fits discussed
in text.\label{fig:figPhi-vs-T-7T-fit }}
\end{figure*}

\subsection{Overview of the phases}

In order to more clearly present our data data we start with a brief
overview of the magnetic structures shown in Fig. \ref{fig:PHDia}.
In the absence of an external magnetic field the low-$T$ magnetic
AF ground state corresponds to a four-sublattices colinear antiferromagnet
(AF) where ferromagnetically ordered plane pairs align antiferromagnetically,
$\uparrow\uparrow\downarrow\downarrow$. With increasing $T$ a series
of six, more complicated, antiferromagnetic phases (AFP$_{n}$) is
found, where one or more, $\uparrow\circ\downarrow$, or, $\downarrow\circ\uparrow$,
3-tuples, containing a paramagentic plane ($\circ$) in the middle,
are inserted between the ferromagentic pairs. This leads to long period
(up to 13 planes) magnetic unit cells. Application of an external
field leads to a ferromagnetic (F) phase through a sequence of metamagnetic
transitions. Below $T\sim12$~K the metamagnetic transitions lead
to two ferrimagnetic phases. The higher-field one, AFF$_{1}$, is
composed from the $\uparrow\uparrow\downarrow$ 3-tuples, while the
lower-field one corresponds to the sequence of the AF unit cells interlaved
with the $\uparrow\uparrow\downarrow$ 3-tuples. Above $T\sim10$~K
five additional ferromagnetic phases (FP$_{n}$ and FP') that contain
paramagentic planes inserted between the ferromagnetic planes appear.
In the FP$_{n}$ phases the paramagnetic planes form pairs, $\circ\circ$.

\subsection{Transient data overview}

In Fig. \ref{fig:figPhi-vs-T-7T-fit } we show selected polarization
rotation transients. A typical transient displays a fast sub picosecond
rise followed by a slower increase on tens-of-picoseconds timescale
and a nanosecond recovery. In the F and AF phases coherent oscillations
are present within the first $\sim10$ ps. At all fields and temperatures
we observe also an initial relatively-small fast transient ($\sim200$-fs)
with the opposite sign to the dominant transient. The transient reflectivity
{[}see Appendix \ref{subsec:Polarization-rotation-due}, Fig. \ref{fig:figAngDep}
b){]} shows a similar shape, but was not studied in detail.

\begin{figure}
\includegraphics[width=0.7\columnwidth]{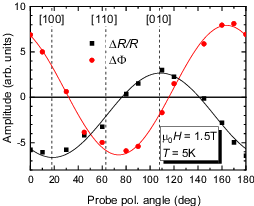}\caption{Angular dependence of the transient reflectivity and polarization
rotation amplitudes (see also Fig. \ref{fig:figAngDep} in Appendix)
in the AFF$_{2}$ phase. The full lines are harmonic fits. The angle
is measured relative to the direction of the magnetic field. The vertical
dashed lines indicate the approximate orientation of the crystal directions.\label{fig:figAvsPhi}}
\end{figure}

The transients show no pump-polarization dependence while the probe
polarization dependence is strong and consistent with the low-$T$
tetragonal symmetry (see also Appendix \ref{subsec:Polarization-rotation-due})
with the $c$-axis and the (sublattice) magnetization along either
the {[}100{]} or {[}010{]} direction.  As shown in Fig. \ref{fig:figAvsPhi}
$\Delta R/R$ shows extrema along the {[}100{]} and {[}010{]} directions
while $\Delta\Phi$ is shifted by $\sim45\lyxmathsym{\textdegree}$
with extrema close to the {[}110{]} direction.

The coherent oscillations are clear and strong in the F phase (see
the low $T$ high field scans in Fig. \ref{fig:figPhi-vs-T-7T-fit }
a). With decreasing magnetic field {[}Fig. \ref{fig:figPhi-vs-T-7T-fit }
(b){]} below $\mu_{0}H\sim4$~T, in the AFF$_{1}$ phase, the coherent
oscillations vanish and the sub-picosecond rise time increases. Decreasing
the field further leads to another qualitative change of the response
below $\mu_{0}H\sim1$ T, in the AF phase, where the response changes
sign and the coherent oscillations reappear, albeit weaker than in
the F phase. The sign change in the AF phase is consistent with rotation
of the tetragonal $c$-axis and sublattice magnetizations around the
surface normal ({[}001{]} pseudo-cubic direction) by $90\lyxmathsym{\textdegree}$.
In other phases no coherent oscillations could be clearly and reproducibly
observed above the noise floor.\footnote{ No data were acquired in the AFP$_{n}$ phases, which remain to be
the subject of further studies.}

\subsection{Analysis}

\begin{figure}[h]
\hfill{}\includegraphics[width=0.9\columnwidth]{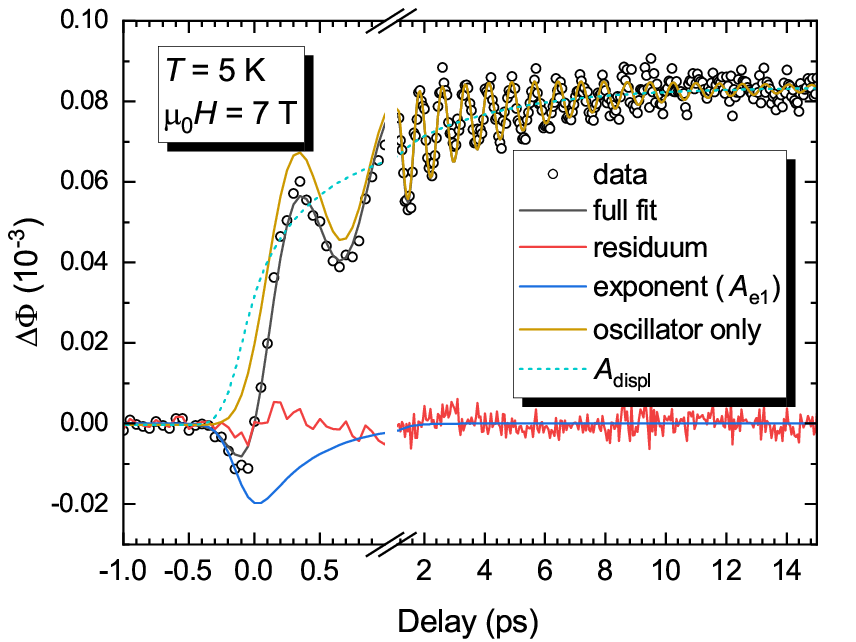}\hspace*{\fill}\caption{DECO fit components in the F phase. The dashed line represents the
instant oscillator equilibrium position. \label{fig:figDPhi-vs-B-fit-comp-7T}}
\end{figure}

\begin{figure}
\hfill{}\includegraphics[width=1\columnwidth]{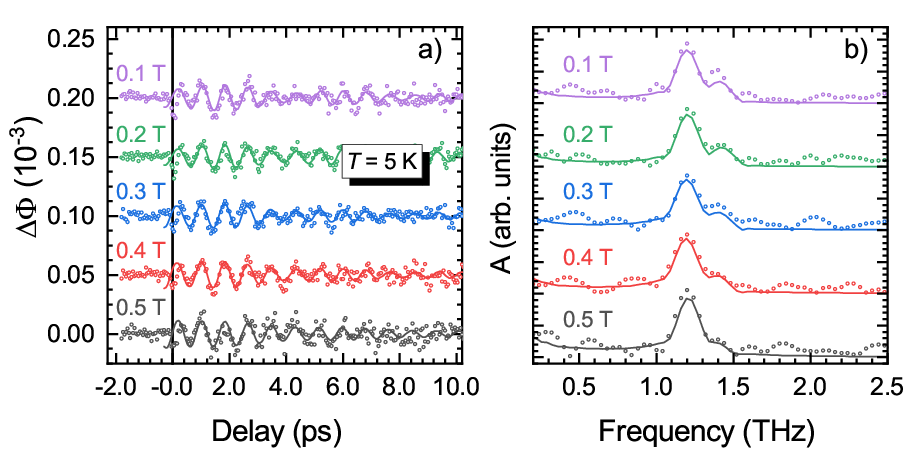}\hspace*{\fill}\caption{Double damped cosine fit to the oscillations in the AF phase (a) and
the corresponding Fourier transforms (b). The solid lines are fits
and the traces are vertically shifted for clarity.\label{fig:Damped-cosine-fit}}
\end{figure}

\begin{figure}
\hfill{}\includegraphics[width=0.9\columnwidth]{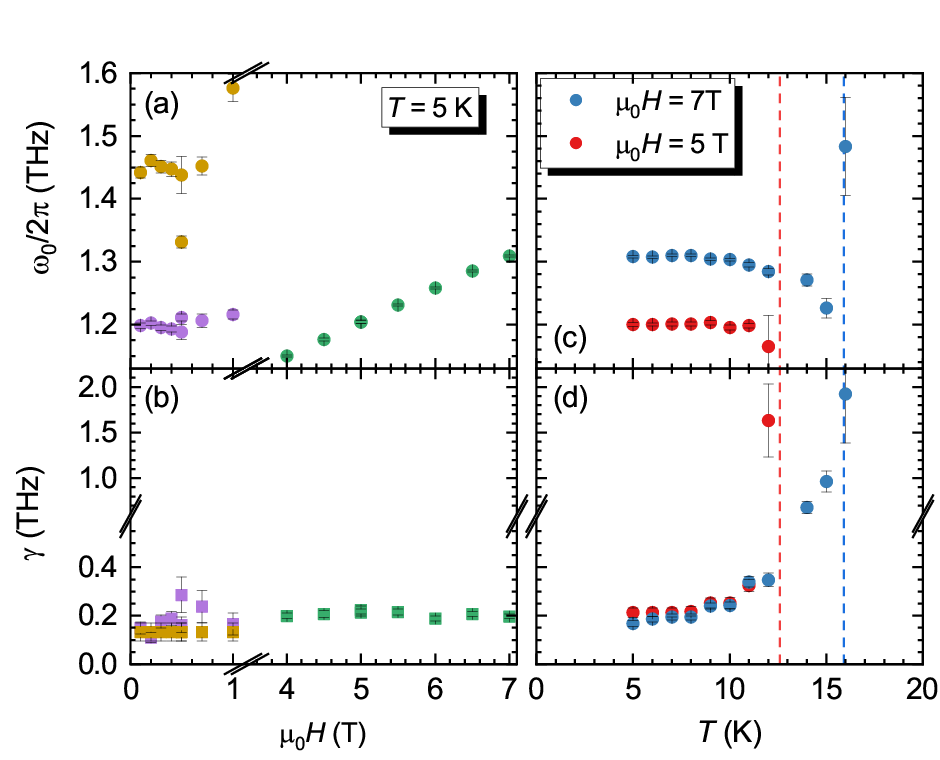}\hspace*{\fill}\caption{Frequency and damping of the coherent oscillations in the AF and F
phases as a function of $H$ (a), (b) and in the F phase as a function
of $T$ (c), (d). The dashed lines represent the F phase boundaries
from \citet{chattopadhyay1994highpressure}.\label{fig:OvsAny}}
\end{figure}

\begin{figure*}
\includegraphics[width=1\textwidth]{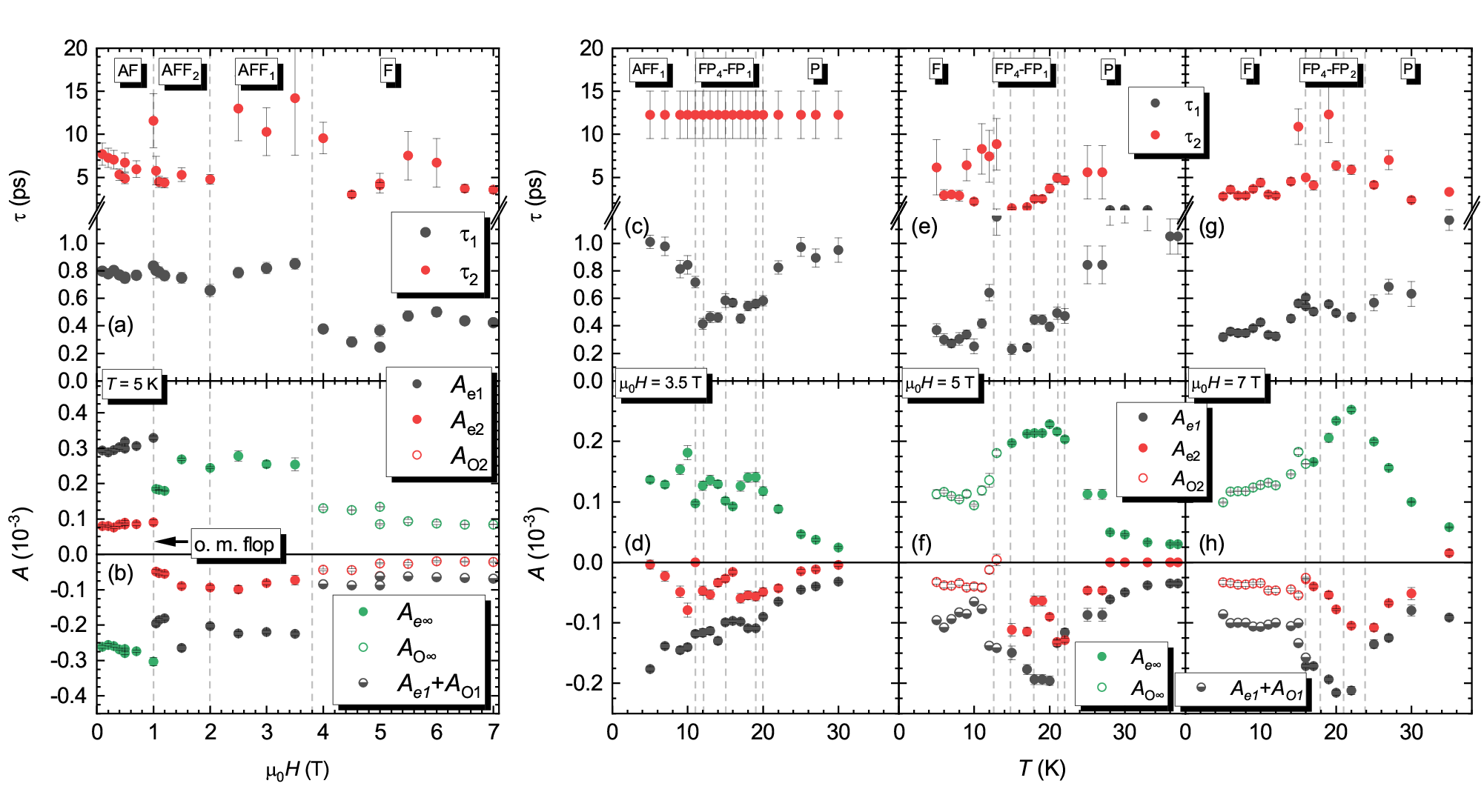}\caption{Magnetic field and temperature dependence of the selected DECO fit
parameters. The open circles correspond to the oscillator while the
full correspond to the exponential components. The half filled symbols
correspond to the total amplitude (see text) of the fast exponential
component in the F phase. The gray dashed lines represent the phase
boundaries from \citet{chattopadhyay1994highpressure}. At the lowest
field {[}$\mu_{0}H=3.5$ T, panels (c) and (d){]} $\tau_{\mathrm{2}}$
was fit globally across all data sets to avoid the fit instabilities
at some individual temperatures.\label{fig:AvsT}}
\end{figure*}

In order to extract some quantitative information from the transient
polarization rotation we use a phenomenological displacive-excitation
of coherent oscillator (DECO) model\citep{zeiger1992theoryfor} fits
to the data in the F phase (see Appendix \ref{subsec:Displacive-excitation-of}
for details), where the coherent oscillations are strong. In other
phases, where the coherent oscillations are weaker or absent, a simpler
three-exponential component model is used.

While one would expect that even in simpler ferromagnets, governed
by the Landau--Lifshitz--Gilbert equation, the dynamics would be
richer than that of a single scalar oscillator, the DECO model can
describe the F-phase coherent oscillations surprisingly well. Due
to the absence of any beating pattern a single displacive coherent
oscillator is assumed. However, in order to fully describe the transients
and the coherent-oscillations phase the displacement, $A_{\mathrm{displ}}$,
that drives the oscillator needs to be modeled by a sum of three pulse-driven
exponential components with different relaxation times,

\begin{equation}
A_{\mathrm{displ}}(t)=\sum_{j}A_{\mathrm{O}j}\int_{0}^{\infty}g(t-u)\exp^{-u/\tau_{j}}du.\label{eq:Adispl}
\end{equation}
Here $g(u)$ corresponds to a normalized Gaussian driving pulse shape,
$A_{\mathrm{O}j}$ the amplitudes and $\tau_{j}$ the relaxation times.
The two exponential components with the amplitudes $A_{\mathrm{O1}}$,
$A_{\mathrm{O2}}$ and relaxation times, $\tau_{\mathrm{1}}\sim0.3-1$
ps, $\tau_{\mathrm{2}}\sim5-20$ ps, respectively, are used to model
the rise time while the third, with a fixed (very slow) relaxation
time, $\tau_{\infty}$, and the amplitude $A_{\mathrm{O\infty}}$,
was used to fit the tens-of-picoseconds plateau (see Fig. \ref{fig:figDPhi-vs-B-fit-comp-7T}).
Since the data were acquired mostly only on short intervals up to
15 ps the actual values of $\tau_{\infty}$ cannot be obtained from
the fits. Limited number of the nanosecond-interval scans indicate
that the actual value of $\tau_{\infty}$ is $\sim0.8$ ns {[}see
inset to \ref{fig:figPhi-vs-T-7T-fit } (b){]}, however, detailed
$T$ and $H$ dependencies were not determined, so a large unbiased
value of $\tau_{\infty}=10$ ns was assumed for the fits.

In general, each of the exponential displacement components can also
contribute to the optical response directly in the same way as to
$A_{\mathrm{displ}}(t)$, but with independent amplitudes, $A_{\mathrm{e}j}$
{[}see Appendix \ref{subsec:Displacive-excitation-of}, Eq. (\ref{eq:PhiDisp})
for details{]}. In the F phase we were able to fit the data by setting
$A_{\mathrm{e}2}$= $A_{\mathrm{e}\infty}=0$, while a finite $A_{\mathrm{e}1}$
was necessary to completely fit the sub 200-fs dynamics. In other
phases all three exponential amplitudes $A_{\mathrm{e1}}$, $A_{\mathrm{e2}}$
and $A_{e\infty}$, corresponding to $\tau_{\mathrm{1}}$, $\tau_{\mathrm{2}}$
and $\tau_{\infty}$, respectively were finite, while all $A_{\mathrm{O}j}$
are set to zero. The selected fits in different phases are shown as
lines in Fig. \ref{fig:figPhi-vs-T-7T-fit }

Moreover, to obtain fair fits around the zero delay the effective
pump-probe pulse cross-correlation width $\tau_{\mathrm{p}}$ was
used as a global free parameter across the given data set to take
into account the initial fast processes that are not included in the
model as a separate component. This reduced the number of fit parameters
and improved the fit convergence with minimal deterioration of the
fit quality.

To analyze the weak coherent oscillations in the AF phase, the three-exponential-fit
residuals were fit once more by a sum of two damped cosine functions,
$\sum A_{i}\exp\left(-\gamma_{i}t\right)\cos\left(\omega_{i}t+\delta_{i}\right)$,
as shown in Fig. \ref{fig:Damped-cosine-fit}. Two oscillatory components
with field independent frequencies and damping are observed in the
AF phase, the stronger at $\sim1.2$ THz and the weaker at $\sim1.4$
THz {[}see Fig. \ref{fig:OvsAny} (a), (b){]}. The weaker one is barely
resolved from the noise, however, the damped cosine fit reliably locks
to the frequency.

The above differences in the fit function components between the F
phase and the other phases, however, do not necessarily imply that
the fundamental physical origin of the nonoscillatory components differs
in different phases. The dynamics of the degree associated with the
oscillator in the F phase might become overdamped/exponential in the
other phases. In such case one or more exponential components would
replace the oscillatory ones. However, if their relaxation times would
not significantly differ from $\tau_{\mathrm{1}}$ and $\tau_{\mathrm{2}}$,
their presence would result only in an apparent renormalization of
the parameters of the three exponential fit. Due to such possibility
we therefore treat at this stage the non-oscillatory amplitudes as
an agnostic metric of the signal, where in the F phase $A_{\mathrm{e1}}+A_{\mathrm{O1}}$,
$A_{\mathrm{O2}}$ and $A_{\mathrm{O3}}$ correspond to $A_{\mathrm{e}1}$,
$A_{\mathrm{e}2}$ and $A_{\mathrm{e}3}$ in the other phases, respectively.

The field and temperature dependence of the frequencies and dampings
obtained from the fits are shown in Fig. \ref{fig:OvsAny}. In the
F phase the frequency shows a linear dependence on $H$ and is virtually
$T$ independent at $\mu_{0}H=5$ T while it shows $\sim8$ \% softening
above $T\sim8$ K at $\mu_{0}H=7$ T. The damping, on the other hand,
increases with increasing $T$ and strongly diverges at the phase
boundary to the FP$_{4}$ phase where the frequency also shows a strong
anomaly. Since the fitting becomes rather unstable at these particular
points the frequency anomaly is probably a fitting artifact.

The field dependence of other relevant fit parameters for the $T=5$
K magnetic field scan is shown in Fig. \ref{fig:AvsT} (a) and (b).
The fast rise time, $\tau_{\mathrm{1}}$, is significantly smaller
in the F state ($\sim0.4$ ps) than in the AF and both AFF phases
($\sim0.8$ ps). It is virtually field independent within each of
the phases. The slower rise time, $\tau_{\mathrm{2}}$, on the other
hand, appears the largest ($\sim12$ ps), but with large error bars,
in the AFF$_{2}$ phase while it is similar ($\sim5$ ps) in the AF,
AFF$_{1}$ and F phases with differences that are comparable to the
data points scatter.

The amplitudes of the fast rise time components, $A_{\mathrm{O1}}$
($A_{\mathrm{e1}}$), and the nanosecond, $A_{\mathrm{O\infty}}$
($A_{\mathrm{e\infty}}$), components {[}see Fig. \ref{fig:AvsT}
(b){]} behave in a similar manner but with opposite signs. The magnitudes
are similar in the AF, AFF$_{1}$ and AFF$_{2}$ phases, but clearly
smaller in the F phase, where in the F phase the sum $A_{\mathrm{e1}}+A_{\mathrm{O1}}$
is considered. The slow-rise-time component amplitude, $A_{\mathrm{O2}}$,
 is significantly smaller than the fast component amplitude $A_{\mathrm{e1}}$
($+A_{\mathrm{O1}}$) in all phases.

The sign change of the transient response and the amplitudes going
from the AFP$_{2}$ to AF phase could be linked to the angular momentum
flop transition of the spins (and concurrently the $[001]_{\mathrm{T}}$
axis) from the {[}100{]} easy axis in the F and AFF$_{n}$ phases
to the {[}010{]} easy axis in the AF phase. The {[}100{]} easy axis
is oriented close to the magnetic field direction and is thus energetically
favorable in the phases with a finite magnetization while the {[}010{]}
 easy axis is at much larger angle with respect to the field and energetically
more favorable in the AF phase.

The temperature dependence of the relevant fit parameters at three
different magnetic fields is shown in Fig. \ref{fig:AvsT} (c)-(h).
The fast rise time, $\tau_{\mathrm{1}}$, in the AFF$_{1}$ phase
shows a drop from $\tau_{\mathrm{1}}\sim1$ ps to $\sim0.7$ ps with
increasing $T$, but remains larger than in the F and FP$_{1}$-FP$_{4}$
phases. In the F phase it is virtually $T$ independent at $\tau_{\mathrm{1}}\sim0.4$
ps. In the FP$_{1}$-FP$_{4}$ phases it has a bit larger value, $\tau_{\mathrm{1}}\sim0.5$
ps, than in the F phase. The two outlier points the $\mu_{0}H=5$
T FP$_{3}$ phase, at $T=12$ K and 13 K, {[}Fig. \ref{fig:AvsT}
(e){]} are most likely a consequence of a fit instability due to the
vanishing coherent oscillations in this region. In the field-magnetized
paramagnetic (P) phase it is scattered around $\sim1$ ps at the lowest
two field magnitudes, while  it continuously rises with increasing
$T$ at the highest field ($\mu_{0}H=7$ T) from the FP$_{2}$ phase
value, $\tau_{\mathrm{1}}\sim0.5$ ps, to an excess of 1 ps at $T=36$
K.

The slower component rise time, $\tau_{2}$, shows quite large scatter,
which is reflected also in the amplitude, $A_{\mathrm{e}2}$, and
no clear distinction between the different phases is observed. At
the lowest field ($\mu_{0}H=3.5$ T) $\tau_{2}$ was fit globally
across all the relevant data sets to avoid fit instabilities at some
individual temperatures.

Turning to the amplitudes, they are systematicaly lower in the F phase
in comparison to the neighbouring phases. With the exception of  $A_{\mathrm{O\infty}}$,
which smoothly connects to the nanosecond exponential component amplitude,
$A_{\mathrm{e\infty}}$ with increasing $T$, at $\mu_{0}H=7$ T {[}Fig.
\ref{fig:AvsT} (h){]}, all component amplitudes show a steep increase
at the F-FP$_{\ensuremath{4}}$ boundary.  A similar, even sharper
steep increase/discontinuity is observed also at the F-AFF$_{1}$
boundary around $\mu_{0}H=4$ T during the field scan at $T=5$ K
{[}see Fig. \ref{fig:AvsT} (b){]}.

\begin{figure}
\hfill{}\includegraphics[width=0.9\columnwidth]{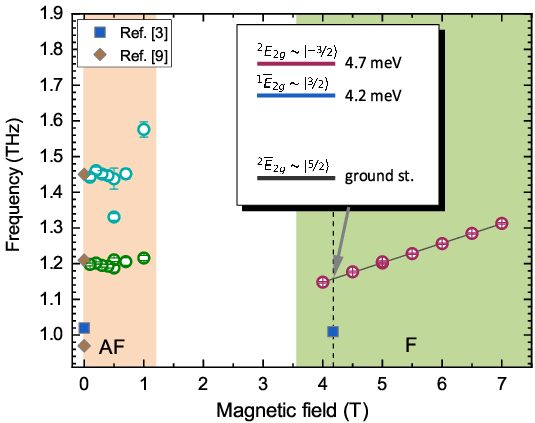}\hspace*{\fill}

\caption{Field dependence of the polarization rotation oscillations frequency
(open symbols) compared to the literature data (full symbols). The
frequencies of oscillations obtained in the antiferromagnetic state
are very close to those obtained in \citep{arai2022multipole}. From
the fit of the frequency field dependence {[}Eq. (\ref{eq:omega}){]} in
the ferromagnetic state (the solid line) we obtain the gyromagnetic
ratio and anisotropy field, using the uni-axial ferromagnetic crystal
model \citep{turov1965physical}. Shematics of the lowest Ce$^{3+}$
4f-derived $\Gamma$-point states in the F phase is shown in the inset.
\label{fig:Freq-comp}}
\end{figure}

\section{Discussion}

\subsection{Coherent oscillations}

To understand the origin of the coherent oscillations we compare the
frequencies obtained using the transient optical spectroscopy to previous
results obtained by means of different techniques. Inelastic neutron
scattering (INS) measurements performed in the F phase at 4.2~K and
4.2~T show one peak at 4.2 meV (1.01 THz) at the \textgreek{G} point.
At these conditions CeSb is in the fully polarized F state with nearly
complete saturation of the magnetic moment. On the basis of the INS
results \citet{halg1986anisotropic} proposed that CeSb can be considered
as a two-level system in the F phase, because the molecular field
dominates over the crystal field and the Ce$^{3+}$ ion 4f-levels
in the ordered state are almost pure Zeeman levels. Their hypothesis
was that the ground state is almost pure $\left|5/2\right\rangle $
and at low $T$ the only possible magnetic excitation is a magnon-like
transverse excitation from $\left|5/2\right\rangle $ to $\left|3/2\right\rangle $
state. At similar conditions, we optically observe a single magnetic
mode, but, at the frequency of 1.14 THz (4.7 meV), which is significantly
(13\%) larger (see Fig. \ref{fig:Freq-comp}).

There are two different reports on excitations in the AF state of
CeSb. The INS measurements done by \citet{halg1986anisotropic} revealed
an excitation with energy of around 4.2 meV (1.01 THz) at the $\Gamma$
point (see Fig. \ref{fig:Freq-comp}) that we have not observed optically.
On the other hand, the Raman scattering measurements reported by \citet{arai2022multipole}
show additional excitations with energies of 5 meV (1.21 THz) and
6 meV (1.45 THz). Both energies coincide with our results. The $6$-meV
excitation was observed also in their INS spectra.

Looking from the point of the Ce$^{3+}$ localized 4f crystal field
level excitations, the ground state in the magnetically ordered states
must be close to the $\left|5/2\right\rangle $ state due to the almost
saturated experimental magnetic moment and originates from the $\Gamma_{8}$
quartet \citep{boucherle1980polarized,takahashi1985anisotropic2}.
The lowest energy excited state, which is observed by inelastic neutron
scattering at $\sim4.2$ meV in AF and F phases, \citep{halg1986anisotropic}
was proposed \citep{takahashi1985anisotropic2} to originate from
the $\Gamma_{7}$ doublet. Totally one expects four more 4f excited
states per Ce$^{3+}$ ion.

For the corresponding collective states at the Brillouin zone $\Gamma$
point the single ion symmetries and splittings should apply in the
P and F phases, while in the AF phase the number of Ce$^{3+}$ ions
in the primitive cell is increased to 4 resulting in 24 states totally.
The local molecular field should split the single ion states, however
some degeneracy persists (see Appendix \ref{sec:Symmetry-considerations})
due to the symmetry equivalence of the crystallographic sites within
the AF tetragonal primitive cell.

In the present experiment we observe two coherent excitations in the
AF phase that coincide with the Raman modes reported by \citet{arai2022multipole}
and should therefore have the same origin. \citeauthor{arai2022multipole}
assigned them to the magnetic excitonic transitions between $\Gamma_{8}$-derived
split states, possibly of vibronic character. The present work indicates
that the frequencies show virtually no magnetic field dependence,
however, this does not rule out the possible magnetic excitonic origin.
Due to the angular momentum flop into the {[}010{]} easy-axis direction
in the AF phase the sublattice magnetizations become nearly perpendicular
to the external magnetic field so the field dependence should be rather
weak.

On the other hand, the BZ back folding can result also in low frequency
optical phonon modes. The phonon dispersions in ReSb \citep{rakshit2008lattice,arai2022multipole}
display a TA phonon branch dispersing up to $\sim1.5$ THz along $\left\langle 00u\right\rangle $
direction. In the AF phase the cubic BZ points $\left(00u\right)$,
where $u=1$ and $\pm1/2$, back fold to the $\Gamma$ point, possibly
producing optical phonons in the $1-1.5$ THz frequency range. This
should in principle be weak for magnetic ordering, however, in CeSb
the magnetic ordering is accompanied with periodic lattice distortion
in the AFP$_{n}$ phases. \citep{mcmorrow1997xrayscattering,iwasa1999evidence}
Moreover, a non-dipersive optical phonon has been reported at 1.7
THz along $\left\langle 0uu\right\rangle $ direction that appears
below $T_{\mathrm{N}}$ in the AF phase. \citep{iwasa2002crystallattice}
The origin of the optically observed AF phase modes therefore remains
unclear.

Due to the small signal to noise ratio and because the impulsive stimulated
Raman scattering (ISRS) coherent-oscillations phase (see Fig. \ref{fig:Damped-cosine-fit})
can be arbitrary in absorbing materials \citep{stevens2002coherent},
it is not possible to distinguish (by means of the fitting) whether
the excitation of these modes in our experiment is via the ISRS or
a displacive mechanism.  The optical Raman transitions between any
of the identical parity electronic states are allowed \citep{kiel1969selection}
in the nonmagnetic tetragonal D$_{4\mathrm{h}}$ symmetry already.
ISRS excitation is therefore symmetry allowed for excitonic transitions
in the ordered phases, which are weakly tetragonal. The absence of
any pump polarization dependence indicates, however, that the coupling
is not a tensor suggesting that a displacive mechanism is more likely.

The displacive excitation can be more symmetry constrained. In bulk
gapless materials the effective displacement is usually a scalar quantity,
since any information about the initial photon polarization and wave
vector is lost after the initial electron/hole pair is inelastically
scattered on a femtosecond timescale. In such case the displacive
excitation can be viewed as a fast Hamiltonian parameter quench. If
such a quench is non adiabatic the state of the system is no longer
the ground state, but a linear combination of the eigenstates of the
perturbed Hamiltonian, which exhibit time evolution. When the parameter
quench preserves the Hamiltonian symmetry the linear combination of
the perturbed eigenstates must transform identically to the original
ground state implying that the ground and displacive excited state
must belong to an identical symmetry-group representation.

In the AF phase, the Ce$^{3+}$-4f-levels derived states at the $\Gamma$
point include (among others) four $\overline{E}_{2\mathrm{g}}$ symmetry
states (see Appendix \ref{sec:Symmetry-considerations}, Tab. \ref{tab:Splitting-of-the}).
Assuming that one corresponds to the ground state there are three
electronic excited states accessible by the displacive excitation. 

Turning to the F phase, the frequency of the single coherent oscillation
virtually linearly depends on the external magnetic field in a manner
that would be expected for the ferromagnetic resonance (FMR) in the
field parallel to an easy axis.\citep{kittel1951ferromagnetic} However,
due to the finite optical penetration depth the oscillations can be
excited only in a $\sim50$ nm thick surface layer\footnote{The optical penetration depth at the 3.1-eV pump photon energy is
$\lambda_{\mathrm{opt}}\sim50$ nm and at the 1.55-eV probe photon
energy $\sim100$ nm as estimated from the optical data in Ref. \citep{kwon1990optical}.} so the demagnetization FMR frequency shift needs to be taken into
account when comparing to the INS results. With the external magnetic
field parallel to the surface and nearly parallel to the {[}100{]}
easy axis direction the Kittel ferromagnetic resonance (FMR) formula
for a thin slab with an in-plane uni-axial anisotropy can be assumed,

\begin{equation}
\omega=\gamma_{\mathrm{M}}\mu_{0}\sqrt{(H+H_{\mathrm{A}})(H+H_{\mathrm{A}}+M_{0})},\label{eq:omega}
\end{equation}
where $H_{\mathrm{A}}$ and $M_{0}$ correspond to the effective anisotropy
field and the F phase magnetization, respectively, while $\gamma_{\mathrm{M}}=g\text{\textmu}_{B}/\hbar$
is the gyromagnetic ratio. Taking an almost saturated Ce$^{3+}$ magnetic
moment of, $\mu_{\mathrm{Ce^{3+}}}\sim2\mu_{\mathrm{B}}$, \citep{halg1986anisotropic}
we get, $\mu_{0}M_{0}\sim0.4$ T, and obtain $g=3.9$ and $\mu_{0}H_{\mathrm{A}}=16.6$~T
from the fit to the data shown in Fig. \ref{fig:Freq-comp}.

The large magnitude of the obtained $g$ factor clearly indicates
that the crystal field effects are not significantly suppressed in
the F phase, contrary to the earlier \citep{halg1986anisotropic}
suggestion, since $g\sim0.9$ would be expected for pure Ce$^{3+}$
Zeeman level splitting. Further, the magnitude of $M_{0}$ is\emph{
too small} \footnote{It can account only for $\sim M_{0}/2H_{0}=1.6\%$ demagnetization
FMR frequency shift.} to explain the 13\% difference between the neutron and optical FMR
frequencies by the shape anisotropy as it can account only for $\sim M_{0}/2H=1.6\%$
demagnetization FMR frequency shift. The excited layer is also relatively
thick in comparison to the atomic scale so any significant addition
surface anisotropy can also be ruled out as the origin of the frequency
difference.

Different sample strain in different experiments can also influence
$H_{\mathrm{A}}$ and lead to different FMR/AFMR frequencies since
the magnetic diagram of CeSb is quite sensitive to the external pressure.
\citep{bartholin1978pressure,chattopadhyay1994highpressure} In our
experiment a mm size sample was glued by GE-varnish on a copper holder
and the surface far from the holder was used so the extrinsic strain
due to the holder/sample thermal expansion mismatch is negligible.
The only possible source of a significant strain could therefore be
built-in strain related to the sample quality and growth conditions.
However, the F-AFF$_{1}$ phase transition, which is pressure sensitive
\citep{bartholin1978pressure}, is observed in our sample at the magnetic
field value of, $\mu_{0}H=3.65$ T {[}see Fig. \ref{fig:figPPExp}
(d){]}, giving no indication that built-in strain would be present.

The observed coherent oscillation therefore \textit{cannot correspond}
to the lowest-energy spin-wave-like branch observed by means of the
INS \citep{halg1986anisotropic} and is assigned to a magnetic excitonic
excitation of another Ce$^{3+}$ 4f-derived collective level. The
group-theoretical symmetry analysis suggests (see Appendix \ref{sec:Symmetry-considerations},
Tab. \ref{tab:Transformation-properties-of}) that the ground state
corresponds to the $\Gamma$-point $^{2}\overline{E}_{2\mathrm{g}}$
($\Gamma_{8}^{+}$) \citep{ye2018extreme} symmetry and can be excited
in a displacive manner (without any external symmetry breaking field)
only to the second $^{2}\overline{E}_{2\mathrm{g}}$ ($\Gamma_{7}^{+}$)
symmetry level. Considering the probe process one should analyze the
Raman tensor for the transitions between these levels, which is of
the A$_{g}$ symmetry,\footnote{When inspecting Tab. \ref{tab:Splitting-of-the} one should take into
account that the matrix element\citep{kiel1969selection} contains
the conjugated final state wave function that can transform differently
in the case of double groups. To obtain the Raman-transition tensor
symmetry between two levels with $^{2}\overline{E}_{2\mathrm{g}}$
symmetry one has to consider the direct product of $^{2}\overline{E}_{2\mathrm{g}}^{*}\otimes{}^{2}\overline{E}_{2\mathrm{g}}={}^{1}\overline{E}_{2\mathrm{g}}\otimes{}^{2}\overline{E}_{2\mathrm{g}}$.} consistent with the observed probe-polarization angular dependence
(see Appendix \ref{subsec:Polarization-rotation-due}, \citet{cracknell1969scattering}
and Fig. \ref{fig:Effect-of-the}).

The $^{2}\overline{E}_{2\mathrm{g}}$ levels are derived from the
linear combinations of the $\left|\nicefrac{-3}{2}\right\rangle $
and $\left|\nicefrac{5}{2}\right\rangle $ Ce$^{3+}$ $^{2}$F$_{5/2}$
ionic states, which  are mixed in the cubic crystal field already.
Since the ground state has almost pure $\left|\nicefrac{5}{2}\right\rangle $
character \citep{halg1986anisotropic} the observed coherent-oscillation
excitation should have dominating $\left|-\nicefrac{3}{2}\right\rangle $
character. The transition $\Delta j_{\mathrm{z}}$ is therefore $\sim4$
resulting in the $g_{\ensuremath{\Delta j_{\mathrm{z}}=1}}\sim1$,
which is close to the ionic \citep{fulde1985magnetic} $g_{J}=6/7$.
Considering the energy of the magnon-like $\left|\nicefrac{3}{2}\right\rangle $-character
excitation \citep{halg1986anisotropic} (see inset to Fig. \ref{fig:Freq-comp})
the level spacings therefore do not appear Zeeman-like so the crystal-field
and quadrupolar bilinear interactions are relatively strong also in
the magnetically ordered phases.

The independence of the pump polarization and the good quality of
the DECP fits indicate that the excitation mechanism is displacive.
On the microscopic level this could be clearly linked to the mixing
of the $\left|\nicefrac{-3}{2}\right\rangle $ and $\left|\nicefrac{5}{2}\right\rangle $
derived $^{2}\overline{E}_{2\mathrm{g}}$ electronic states. The mixing
should depend on the quasiparticle distribution function in the p-f
hybridized conduction bands, which is, upon the photo-excitation,
changed on a subpicosecond scale. This corresponds to a nonadiabatic
effective Hamiltonian quench, which leads to a change of the $\left|\nicefrac{-3}{2}\right\rangle $-
and $\left|\nicefrac{5}{2}\right\rangle $- derived wavefunctions
mixing. As a result, a \textit{quantum superposition state} of the
$\left|-\nicefrac{3}{2}\right\rangle $- and $\left|\nicefrac{5}{2}\right\rangle $-dominated
$^{2}\overline{E}_{2\mathrm{g}}$ wavefunctions, with a well defined
quantum phase, is created. The observed coherent oscillation therefore
corresponds to the \textit{real-time quantum evolution} and dephasing
of the superposition state. 

On the macroscopic level, the observable couples to the quadrupolar
degrees of freedom according to the probe polarization angular dependence
of the signal (see also Appendix \ref{subsec:Polarization-rotation-due}).
The oscillation therefore could be associated with a direct coherent
collective excitation of a (secondary) quadrupolar order parameter
that is adjoined \citep{morin1990chapter} to the dipolar order parameter
in the unfilled 4f shell systems. Based on the symmetry (Appendix
\ref{subsec:Polarization-rotation-due}) such excitation couples to
to the longitudinal modulation of the dipolar (magnetization) order
parameter.

The absence of the coherent oscillations due to the lowest energy
transition to the collective transverse spin-wave-like state with
the $^{1}\overline{E}_{2\mathrm{g}}$ ($\left|\nicefrac{3}{2}\right\rangle $)
character observed by INS \citep{halg1986anisotropic} is somewhat
surprising. The displacive excitation is symmetry allowed due to the
external symmetry-breaking magnetic field,\footnote{In Ni, for example, spin waves can be displacively excited\citep{vankampen2002alloptical}
when the external magnetic field is applied at some angle to the anisotropy
field.} which is at a finite angle with respect to the {[}100{]} axis. However,
due to the tetragonal symmetry the transverse magnetization components
couple quadraticaly to the reflectivity in the lowest order (see Appendix
\ref{subsec:Polarization-rotation-due}) and are likely below the
detection limit.

Reasons for the absence of any coherent modes in the FP$_{n}$ and
AFF$_{n}$ magnetic phases are unclear. There is no fundamental account
prohibiting their coupling to the optical probe and the non-oscillating
components  are qualitatively similar in all phases, suggesting that
the same degrees of freedom are involved. A possible reason could
be a larger dephasing. Different intrinsic dephasing (on the quantum
level) due to different electron scattering in different phases \citep{ye2018extreme}
would be plausible, since is clearly different in different phases\citep{aoki1985fermisurface,kaneta2000theoretical}.
On the other hand,  stronger disorder, which contributes to the dephasing,
could also be expected in the intermediate phases due to their larger
magnetic unit cells and similar free energies. As a result, long magneto-structural
coherence lengths are unlikely, especially in the presence of the
pulsed laser perturbation.

\subsection{Non-oscillatory response}

Turning to the non-oscillatory relaxation components we observe a
qualitatively similar response in all phases. The probe polarization
dependence and the presence of the components in the external-field-magnetized
P phase suggests that it could be linked (on a phenomenological level)
to the longitudinal transient sublattice magnetization which is strongly
coupled to the electronic structure due to the p-f mixing. \citep{aoki1985fermisurface,kaneta2000theoretical}

The dynamics is driven by the non equilibrium carrier redistribution
and their excess energy transfer to the lattice. Due to the absence
of the gap it is not surprising that the dominant relaxation is happening
on a (sub-) picosecond timescale. The difference of this timescale
in different phases can be attributed to their electronic structures.
\citep{kaneta2000theoretical} The similar $\tau_{1}\sim1$ ps in
the AF, AFF$_{1}$ and AFF$_{2}$ is suggests that their electronic
structure is similar, while significantly shorter $\tau_{1}\sim0.5$
ps in the FP$_{1}$-4 phases indicates that their electronic structure
is more similar to the F phase. The minor and slower signal-rise component
timescale could also be of the related origin involving different
parts of the Fermi surface. However, in order to obtain more detailed
insight, kinetics models of the relaxation would need to be solved,
which is beyond the sope of the present work.

\section{Summary and conclusions}

We performed time-resolved measurements of the magneto-optic response
in CeSb with systematic variation of the temperature and magnetic
field to obtain the transient magneto-optic response in different
magnetically ordered phases.

In the AF and F phases we observe damped coherent oscillations of
the magneto-optic response. In the fully polarized F phase, a single
coherent oscillation is observed with the spectroscopic g-factor,
$g=3.94$. The observed frequency differs by 13\% from the \textgreek{G}-point
inelastic-neutron-scattering observed magnetic excitation. \citep{halg1986anisotropic}
The oscillation is attributed to the \textit{real-time coherent quantum
evolution} of the displacively induced \textit{quantum superposition}
of the magnetic excitonic  Ce$^{3+}$-4f-derived collective states.
To our best knowledge such  real-time coherent evolution of a quantum
collective state has not ben reported before. The finding opens the
way for posibility to control such excitonic collective states on
the quntum level.

The coherent oscillation frequency and the magnitude of the $g$ factor
together with the detailed group-theoretical symmetry analysis put
the $\left|\nicefrac{-3}{2}\right\rangle $-character $^{2}\overline{E}_{2\mathrm{g}}$
level in the vicinity of the $\left|\nicefrac{3}{2}\right\rangle $-character
$^{1}\overline{E}_{2\mathrm{g}}$level. This indicates that the level
spacing is, contrary to the literature,\citep{halg1986anisotropic}
not Zeeman-like and the crystal-field and quadrupolar bilinear interactions
are relatively strong also in the magnetically ordered phases.

In the AF phase two weaker coherent oscillations with \textit{magnetic
field independent} frequencies of 1.20 THz and 1.45 THz are observed.
The frequencies correspond to the modes recently observed also by
means of Raman scattering {[}14{]}. While it is likely that the modes
have the magnetic excitonic origin, as in the F phase, the lattice
origin cannot be entirely ruled out.

The non-oscillatory part of the transients shows faster sub-picosecond
dynamics in the F and FP$_{1}$-FP$_{4}$ phases in comparison to
the AF, AFF$_{1}$ and AFF$_{2}$ phases. This suggests that the electronic
structures, which affect the kinetics of the hot electron energy relaxation,
in the F and FP$_{n}$ differ from the electronic structures in the
the AF, AFF$_{1}$ and AFF$_{2}$ phases.
\begin{acknowledgments}
The authors acknowledge the financial support of Slovenian Research
and Innovation Agency (research core funding No-P1-0040 and young
researcher funding No. 50504). We would also like to thank V. V. Kabanov
and A. Shumilin for fruitful discussions.
\end{acknowledgments}

\appendix 

\section{}

\section{Pump fluence dependence of optical transients}

\begin{figure}
\includegraphics[width=0.9\columnwidth]{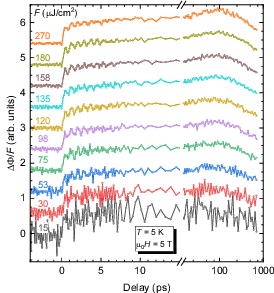}\caption{Fluence dependence of the normalized transient polarization rotation
at $T=5$ K and $\mu_{0}H=5$ T. Note the logarithmic scale after
the break.\label{fig:Fdep}}
\end{figure}

In Fig. \ref{fig:Fdep} we show the pump fluence dependence of $\Delta\Phi$.
The fluence normalized scans are virtually $F$ independent indicating
a linear scaling of the transients with increasing $F$.

\section{Displacive excitation of coherent oscillations\label{subsec:Displacive-excitation-of}}

Following the standard derivation \citep{zeiger1992theoryfor} for
displacive driven oscillator we assume a single harmonic oscillator
driven by a displacement composed from three pulse-driven exponential
components (\ref{eq:Adispl}) and obtain,\begin{widetext}

\begin{alignat}{1}
\Delta\Phi(t)= & \sum\left[(A_{\mathrm{e}j}+A_{\mathrm{O}j})\int_{0}^{\infty}G(t-u)\exp^{-u/\tau_{j}}du\right.-\label{eq:PhiDisp}\\
 & \left.A_{\mathrm{O}j}\int_{0}^{\infty}G(t-u)\exp^{-\gamma u}[\cos(\Omega u)-\beta_{j}\sin(\Omega u)]du\right],\nonumber 
\end{alignat}
\end{widetext}where, 
\begin{align}
\beta_{j}=(1/\tau_{j}-\gamma)/\Omega
\end{align}
and 
\begin{equation}
G(t)=\sqrt{\frac{2}{\pi}}\frac{1}{\tau_{\textrm{p}}}e^{-\frac{2t}{\tau_{\textrm{p}}^{2}}}.\label{Gauss}
\end{equation}
The amplitudes of the exponential displacement components with relaxation
times, $\tau_{j}$, are given by $A_{Oj}$, while $\Omega$ and $\gamma$
correspond to the oscillator (renormalized) frequency and damping,
respectively. $\tau_{\textrm{p}}$ corresponds to the effective pump-probe
pulse cross-correlation width. $A_{\mathrm{e}j}$ allow for the possibility
that any of the exponential displacement components contributes also
directly to the optical response.

\section{Polarization rotation due to Cotton-Mouton effect in the cubic symmetry\label{subsec:Polarization-rotation-due}}

The second order magneto-optical tensor describing the magnetic linear
dichroism and birefringence in $Fm\overline{3}m$ symmetry is given
by, \citep{gallego2019automatic}\begin{widetext}

\begin{equation}
G=\left[\begin{array}{ccc}
c_{11}M_{x}^{2}+c_{21}\left(M_{y}^{2}+M_{z}^{2}\right) & c_{55}M_{x}M_{y} & c_{55}M_{x}M_{z}\\
c_{55}M_{x}M_{y} & c_{11}M_{y}^{2}+c_{21}\left(M_{x}^{2}+M_{z}^{2}\right) & c_{55}M_{y}M_{z}\\
c_{55}M_{x}M_{z} & c_{55}M_{y}M_{z} & c_{11}M_{z}^{2}+c_{21}\left(M_{x}^{2}+M_{y}^{2}\right)
\end{array}\right],
\end{equation}
\end{widetext}where $c_{ij}$ correspond to the three independent
coefficients and $M_{j}$ to the magnetization components. Assuming
the static magnetization, $M_{0}$, along {[}100{]} the tensor can
be linearized with respect to the transient magnetization, $\left(\Delta M_{x},\Delta M_{y},\Delta M_{z}\right),$\begin{widetext}

\begin{equation}
G\thickapprox\left[\begin{array}{ccc}
c_{11}\left(M_{0}^{2}+2M_{0}\Delta M_{x}\right) & c_{55}M_{0}\Delta M_{y} & c_{55}M_{0}\Delta M_{z}\\
c_{55}M_{0}\Delta M_{y} & c_{21}\left(M_{0}^{2}+2M_{0}\Delta M_{x}\right) & 0\\
c_{55}M_{0}\Delta M_{z} & 0 & c_{21}\left(M_{0}^{2}+2M_{0}\Delta M_{x}\right)
\end{array}\right].
\end{equation}
\end{widetext}The off-diagonal components couple to the transverse
transient magnetization while the diagonal to the longitudinal one.

Constraining the light propagation along the {[}001{]} direction one
obtains,
\begin{eqnarray}
\Delta R/R & = & \left[a+b\cos(2\phi)\right]M_{0}\Delta M_{x}+o(\Delta M_{x}^{2},\Delta M_{y}^{2}),\nonumber \\
\Delta\Phi & = & b\sin(2\phi)M_{0}\Delta M_{x}+o(\Delta M_{x}^{2},\Delta M_{y}^{2}),\label{eq:transint-resp}
\end{eqnarray}
where $\phi$ corresponds to the probe polarization angle with respect
to the {[}100{]} direction. $a$ and $b$ are rational polynomial
functions of the real, $n,$ and imaginary part, $k$, of the probe
refraction index as well as the magneto-optical tensor factors $c_{11}$
and $c_{21}$. Unless, $\left|c_{55}\Delta M_{y}\right|>\left|\left(c_{21}-c_{11}\right)M_{0}\right|$,
the dependence on $\Delta M_{y}$ is quadratic. Both, the transient
reflectivity and polarization rotation couple to the longitudinal
transient magnetization, $\Delta M_{x}$, only. The behavior is similar
also in the tetragonal case when both, the tetragonal axis and $M_{0}$,
are oriented along the pseudo-cubic {[}100{]} or {[}010{]} directions.

The transverse transient magnetization along the light propagation
direction, $\Delta M_{z}$, can be detected through the polar MOKE
at any probe polarization if the coupling constant is large enough.
The coupling is linear and the probe polarization independent only,
when $\left|\alpha\Delta M_{z}\right|>\left|\left(c_{21}-c_{11}\right)M_{0}\right|$,
with $\alpha M_{z}$ corresponding to the component of the magnetooptical
gyration\citep{zvezdin1997modernmagnetooptics} vector.

\begin{figure}
\bigskip{}
\includegraphics[width=1\columnwidth]{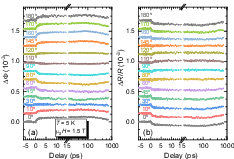}\caption{Angular dependence of (a) polarization rotation and (b) reflectivity
transients at $T=5$ K and $\mu_{0}H=1.5$ T, in the AFF$_{2}$ phase.
Note the logarithmic scale after the breaks.\label{fig:figAngDep}}
\end{figure}

\begin{figure}
\includegraphics[width=0.8\columnwidth]{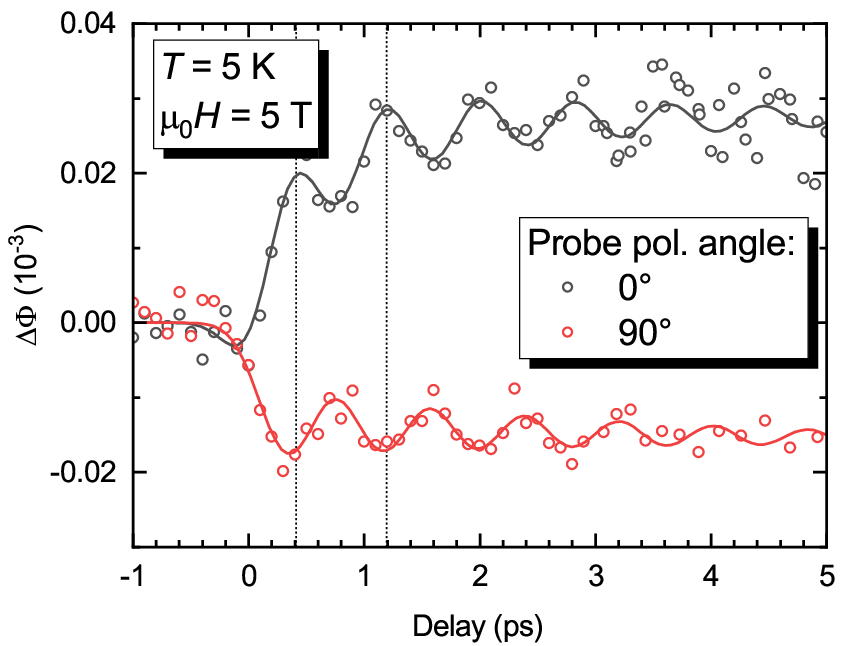}\caption{Effect of the probe polarization on the polarization rotation in the
F phase. The solid lines are the DECO (\ref{eq:PhiDisp}) fits.\label{fig:Effect-of-the}}
\end{figure}

In our case (see Figs. \ref{fig:figAvsPhi}, \ref{fig:figAngDep}
and \ref{fig:Effect-of-the}) the experimental extrema of the transient
reflectivity amplitude are for probe polarizations along {[}100{]}
and {[}010{]} crystal directions while the transient polarization
rotation is shifted by $\sim50{}^{\circ}$ with respect to the transient
reflectivity. Both signals are therefore dominated by the modulation
of the diagonal dielectric tensor components that couple (on the macroscopic
level) to the the longitudinal transient magnetization as well as
the orthorhombic anisotropy of the electronic system.

\section{Symmetry considerations\label{sec:Symmetry-considerations}}

In the paramagnetic phase ($Fm\overline{3}m$) the Ce$^{3+}$ ion
4f $J=5/2$ multiplet derived states correspond to two dimensional
$\overline{E}_{2g}$ ($\Gamma_{7}^{+})$ and four dimensional $\overline{F}_{g}$
($\Gamma_{8}^{+})$ representations of the $\Gamma$ point group ($m\overline{3}m$).
Assuming the highest possible symmetry magnetic groups compatible
with the experimental tetragonal symmetry \citep{hulliger1975lowtemperature}
we obtain \citep{perez-mato2015symmetrybased} $I4/mm'm'$ (N. 139.537)
and $P_{c}4/ncc$ (N. 130.432) magnetic groups for the F and AF state,
respectively. In the F phase \footnote{In an external magnetic field along a 4 fold axis in the P phase as
well.} the $\Gamma$-point degeneracy is completely removed while in the
AF phase the states remain doubly degenerate (see Tab. \ref{tab:Splitting-of-the}).
\citep{elcoro2020application,cracknell1968crystal} Moreover due to
the Brillouin zone folding \citep{jang2019directvisualization} additional
18 back-folded 4f-derived bands appear at the $\Gamma$ point in the
AF phase.

Looking at the direct products of the involved corepresentations\citep{cracknell1968crystal}
(see Tab. \ref{tab:Splitting-of-the}), all transitions are Raman
allowed. %

\begin{table*}
\begin{tabular}{lcccccc}
\noalign{\vskip5pt}
Phase & \hspace*{4pt} & P & \hspace*{4pt} & F & \hspace*{4pt} & AF\tabularnewline[5pt]
\hline 
\hline 
\noalign{\vskip5pt}
Magnetic space group &  & $Fm\overline{3}m$ &  & $I4/mm'm'$ &  & $P_{c}4/ncc$\tabularnewline[5pt]
\noalign{\vskip5pt}
Point group at $\Gamma$ &  & $m\overline{3}m$ &  & $4/mm'm'$ &  & $4/mmm1'$\tabularnewline[5pt]
\noalign{\vskip5pt}
Ce$^{3+}$ site symmetry &  & $m\overline{3}m$ (4a) &  & $4/mm'm'$ (2a) &  & $4m'm'$ (4c)\tabularnewline[5pt]
\hline 
\noalign{\vskip5pt}
\multirow{2}{*}{\vtop{\hbox{\strut Induced $J=5/2$ multiplet}\hbox{\strut  irreducible
band corepresentations}}} &  & $\overline{E}_{2\mathrm{g}}$ ($\Gamma_{7}^{+})$ &  & $^{1}\overline{E}_{2\mathrm{g}}\oplus^{2}\overline{E}_{2\mathrm{g}}$ &  & \multirow{1}{*}{$2\overline{E}_{2\mathrm{g}}\oplus2\overline{E}_{2\mathrm{u}}$}\tabularnewline[5pt]
\noalign{\vskip5pt}
 &  & $\overline{F}_{\mathrm{g}}$ ($\Gamma_{8}^{+})$ &  & $^{1}\overline{E}_{1\mathrm{g}}\oplus^{2}\overline{E}_{1\mathrm{g}}\oplus^{1}\overline{E}_{2\mathrm{g}}\oplus^{2}\overline{E}_{2\mathrm{g}}$ &  & $2\overline{E}_{1\mathrm{g}}\oplus2\overline{E}_{2\mathrm{g}}\oplus2\overline{E}_{1\mathrm{u}}\oplus2\overline{E}_{2\mathrm{u}}$\tabularnewline[5pt]
\hline 
\noalign{\vskip5pt}
\multicolumn{7}{l}{\medskip{}
}\tabularnewline[5pt]
\end{tabular}\\
\setlength{\extrarowheight}{3pt}%
\begin{tabular}{c|cccc}
$4/mm'm'$ (F) & $^{1}\overline{E}_{1\mathrm{g}}$ & $^{2}\overline{E}_{1\mathrm{g}}$ & $^{1}\overline{E}_{2\mathrm{g}}$ & $^{2}\overline{E}_{2\mathrm{g}}$\tabularnewline
\hline 
$^{1}\overline{E}_{1\mathrm{g}}$ & $^{2}E_{g}$ & $A_{\mathrm{g}}$ & $^{1}E_{\mathrm{g}}$ & $B_{\mathrm{g}}$\tabularnewline
$^{2}\overline{E}_{1\mathrm{g}}$ & $A_{\mathrm{g}}$ & $^{1}E_{\mathrm{g}}$ & $B_{\mathrm{g}}$ & $^{2}E_{g}$\tabularnewline
$^{1}\overline{E}_{2\mathrm{g}}$ & $^{1}E_{\mathrm{g}}$ & $B_{\mathrm{g}}$ & $^{2}E_{g}$ & $A_{\mathrm{g}}$\tabularnewline
$^{2}\overline{E}_{2\mathrm{g}}$ & $B_{\mathrm{g}}$ & $^{2}E_{g}$ & $A_{\mathrm{g}}$ & $E_{1\mathrm{g}}$\tabularnewline
\end{tabular}\hspace{0.5cm}%
\begin{tabular}{c|cc}
$4/mmm1'$ (AF) & $\overline{E}_{1\mathrm{g}}$ & $\overline{E}_{2\mathrm{g}}$\tabularnewline
\hline 
$\overline{E}_{1\mathrm{g}}$ & $A_{1\mathrm{g}}+A_{2\mathrm{g}}+E_{\mathrm{g}}$ & $B_{1\mathrm{g}}+B_{2\mathrm{g}}+E_{\mathrm{g}}$\tabularnewline
$\overline{E}_{2\mathrm{g}}$ & $B_{1\mathrm{g}}+B_{2\mathrm{g}}+E_{\mathrm{g}}$ & $A_{1\mathrm{g}}+A_{2\mathrm{g}}+E_{\mathrm{g}}$\tabularnewline
\end{tabular}

\caption{Irreducible corepresentations of the Ce$^{3+}$ ion $J=5/2$ 4f ($^{2}$F$_{5/2}$)
multiplet derived states at the Brillouin zone $\Gamma$ point and
direct product decompositions of the relevant corepresentations.\label{tab:Splitting-of-the}}
\end{table*}

\begin{table}
\setlength{\extrarowheight}{3pt}%
\begin{tabular}{c|c}
correp. & base f.\tabularnewline
\hline 
$^{1}\overline{E}_{1\mathrm{g}}$ & $\left|-\nicefrac{1}{2}\right\rangle $\tabularnewline
$^{2}\overline{E}_{1\mathrm{g}}$ & $\left|\nicefrac{1}{2}\right\rangle $\tabularnewline
$^{1}\overline{E}_{2\mathrm{g}}$ & $\left|-\nicefrac{5}{2}\right\rangle $,$\left|\nicefrac{3}{2}\right\rangle $\tabularnewline
$^{2}\overline{E}_{2\mathrm{g}}$ & $\left|\nicefrac{5}{2}\right\rangle $,$\left|-\nicefrac{3}{2}\right\rangle $\tabularnewline
\end{tabular}

\caption{Transformation properties of the $J=5/2$ multiplet base functions
in the F-state magnetic point group $4/mm'm'$.\label{tab:Transformation-properties-of}}
\end{table}

\begin{widetext}
\clearpage
\end{widetext}

\bibliographystyle{apsrev4-2}
\bibliography{biblio}

\end{document}